\documentclass{article}

\usepackage{PRIMEarxiv}

\usepackage[utf8]{inputenc} 
\usepackage[T1]{fontenc}    
\usepackage{hyperref}       
\usepackage{url}            
\usepackage{booktabs}       
\usepackage{amsfonts}       
\usepackage{nicefrac}       
\usepackage{microtype}      
\usepackage{lipsum}
\usepackage{fancyhdr}       
\usepackage{graphicx}       
\graphicspath{{media/}}     

\title{Evaluating Utility of Memory Efficient Medical Image Generation: A Study on Lung Nodule Segmentation}

\author{
  Kathrin Khadra \\
  RYVER.AI, Munich, Germany \\
  \texttt{kathrin.khadra@ryver.ai} \\
   \AND
     Utku Türkbey \\
  RYVER.AI, Munich, Germany \\
  \texttt{utku.turkbey@ryver.ai} \\
}

\begin{document}
\maketitle

\begin{abstract}
The scarcity of publicly available medical imaging data limits the development of effective AI models. This work proposes a memory-efficient patch-wise denoising diffusion probabilistic model (DDPM) for generating synthetic medical images, focusing on CT scans with lung nodules. Our approach generates high-utility synthetic images with nodule segmentation while efficiently managing memory constraints, enabling the creation of training datasets. We evaluate the method in two scenarios: training a segmentation model exclusively on synthetic data, and augmenting real-world training data with synthetic images. In the first case, models trained solely on synthetic data achieve Dice scores comparable to those trained on real-world data benchmarks. In the second case, augmenting real-world data with synthetic images significantly improves segmentation performance. The generated images demonstrate their potential to enhance medical image datasets in scenarios with limited real-world data.
\end{abstract}


\section{Introduction}

Accessing radiology data to train deep learning models can be a challenge. The reasons can be manifold, often privacy laws prevent people from accessing patient information, sometimes the quality of the data is not good enough or the data is scarce in general. One possible solution could be synthetic data. Specifically, to use synthetic data as training data together with real-world data or as a full replacement for real-world data. In recent years, papers have generated medical images using various methods. Multiple researchers have used Generative Adversarial Networks (GANs) and Variational Autoencoders (VAEs) \cite{2021ganreview, 2024ganreview, genreview}. Kwon et al. \cite{braingan} created a hybrid of a VAE and a GAN, the $\alpha$-GAN to generate 3D brain images from noise.  A StyleGAN, which enables manipulating the features of an image, was used for medical images by Hong et al. \cite{stylegan}. Thambawita et al. \cite{singanseg} have generated images with the corresponding segmentation mask using a GAN approach and trained a segmentation model using synthetic and real-world data. The authors could show that the synthetic data performs similarly to the real-world data and can improve the segmentation model when a substantial amount of real data is missing. 

Besides GANs and VAEs, diffusion models have gained more and more attention over the last years \cite{genreview, diffmodelreview}. Pinaya et al. \cite{pinaya2022brain} utilizes a latent diffusion model to generate Brain MRIs. This saves memory compared to working in the pixel space directly, as the model operates on a compressed latent space. Latent diffusion models were also used by Zhu et al. \cite{makevolume}. The model was trained to create 3D volumes out of 2D slices in an image-to-image translation approach, instead of creating 3D volumes directly. Sagers et al. \cite{skinclassifier} generated images of skin diseases with methods based on a latent model. The generated images augmented the real-world data in a malignancy classification task and improved the accuracy of the model. Two segmentation models were trained on real and synthetic polyp images both achieving a similar intersection over unit by Macháček et al. \cite{masklatent}. The synthetic data was generated with a latent model which was conditioned on segmentation masks. To create large histopathology patches, Aversa et al. \cite{histo} developed a new patch-wise sampling method that uses a latent diffusion model. This way limited memory can be overcome by subsequently generating parts of the image instead of the whole image from scratch.

On the other hand, Friedrich et al. \cite{wavelet} created wavelets out of 3D medical images and trained a Denoising Diffusion Probabilistic Model (DDPM)  on them, using wavelets reduces the spatial dimension of the data significantly. With this, they reach a resolution of $256$ cubic on state-of-the-art hardware, which, to our knowledge, is the highest resolution reached so far for 3D imaging modalities. Guiding the DDPM with a segmentation mask, to create images according to the specific segmentations, was utilized by Konz et al. \cite{segguided}.  Furthermore, Khader et al. \cite{firas} synthesized various medical images with a DDPM and pre-trained a breast segmentation algorithm with them. The final segmentation model presented a higher Dice score with the self-supervised pre-training through synthetic data.
Besides image generation and using the images as training data, diffusion models were also used to create segmentations or detect anomalies directly \cite{segjulia, medsegdif, anomalyjulia, anomalyWyatt, patchddm}.

In this work, we investigate the memory-efficient generation of medical images using a patch-wise DDPM inspired by Bieder et al. \cite{patchddm}. Their approach trains a DDPM on patches to create segmentation masks of these images. In this work, patch-wise training is used to synthesize images memory-efficiently. Additionally, we condition the DDPM with a segmentation mask similar to \cite{masklatent}. Additionally, we investigate the utility of the synthetic data. For this, we compare the following segmentation models against a segmentation model trained on only real-world data: segmentation model trained on only the synthetic data,  segmentation model trained on real-world and synthetic data that augments data points with a low Dice score (DSC). While medical images generated by GANs and VAEs have been tested on the downstream tasks quite frequently, it is still less common for diffusion models \cite{2021ganreview, 2024ganreview,genreview, diffmodelreview}. Furthermore, the diffusion model papers that test on a downstream task have not yet both trained a downstream task on only synthetic data and augmented the real-world dataset. Thus, the novelty of this work lies in first synthesizing images with the patch-wise DDPM approach of Bieder et al. \cite{patchddm} and a more thorough analysis of the effect the synthetic data has on a downstream task.  

\section{Materials and Methodology}

\subsection{Lung Image Database Consortium and Image Database Resource Initiative Dataset (LIDC-IDRI)}

For this work, we utilize the Lung Image Database Consortium and Image Database Resource Initiative (LIDC-IDRI) dataset, a public lung nodule analysis and computer-aided diagnosis dataset \cite{lidc-article, lidc-data}. It contains diagnostic and lung cancer screening thoracic CT scans collected from 1018 cases. Each scan is reviewed by four experienced radiologists, who provide detailed annotations of lung nodules, including their likelihood of malignancy \cite{lidc-article, lidc-data}. This multi-reader approach ensures a comprehensive and nuanced dataset, allowing for variability in interpretation, which is reflective of real-world clinical settings \cite{lidc-article}.

The nodules in the LIDC-IDRI dataset are classified into three categories based on their size \cite{lidc-article, lidc-data}:

\begin{itemize}
\item Nodules $\geq$ 3mm: These nodules are clinically significant as they may indicate potential malignancies.
\item  Nodules $<$ 3mm: These nodules are often considered benign.
\item Non-nodule annotations: These include other pulmonary lesions that are not considered nodules by clinical experts.
\end{itemize}

In this study, we focus exclusively on nodules larger than 3mm, as nodules above this threshold are of greater clinical interest and more relevant for early-stage lung cancer detection \cite{lidc-article}. This selection allows us to concentrate on the generation of synthetic data that reflects the most clinically actionable nodules. The LIDC-IDRI dataset provides approximately 2669 nodules, marked bigger than 3mm by at least one radiologist.

\subsection{Denoising Diffusion Probabilistic Models (DDPMs)}

Denoising Diffusion Probabilistic Models (DDPMs) are a kind of generative models that have gained attention for their ability to generate high-quality data, such as images, through a stochastic diffusion process \cite{ddpm,diffmodel_gan}. Unlike other generative models like GANs, DDPMs rely on a probabilistic forward and reverse diffusion process, which gradually adds and removes noise to and from the data \cite{ddpm,ddpm_improved,diffmodel_gan}.

\subsubsection{Forward Diffusion Process}

The DDPM process starts with a forward diffusion that incrementally noises the data with Gaussian noise over a series of timesteps \cite{ddpm,ddpm_improved}. This process transforms an input data sample, such as a lung CT scan, into pure Gaussian noise after a sufficient number of steps. Let the original data sample be denoted as $x_0$, and let $x_t$ represent the data after applying $t$ noise steps. In each step, noise is added according to a variance schedule, following the equation \cite{ddpm,ddpm_improved}:

\begin{equation}
    q (x_t|x_{t-1})=\mathcal{N}(x_t;\sqrt{\alpha_t}x_{t-1},(1-\alpha_t)\mathbb{I})
\end{equation}

where $\alpha_t$ is a time-dependent scaling factor, and the noise is sampled from a Gaussian distribution. As $t$ increases, $x_t$ becomes progressively noisier, until at the final timestep $T$, the data is approximately drawn from a standard Gaussian distribution.

\subsubsection{Reverse Diffusion Process}

The DDPM is supposed to learn the reverse diffusion process, which progressively denoises the noisy data, recovering the original distribution of the data \cite{ddpm,ddpm_improved}. This reverse process is modeled as a series of conditional Gaussian transitions according to the following equation \cite{ddpm,ddpm_improved}:

\begin{equation}
    p_\theta (x_{t-1}|x_{t})=\mathcal{N}(x_{t-1};\mu_\theta(x_t,t),\Sigma_\theta(x_t,t))
\end{equation}

where $\mu_\theta$ and $\Sigma_\theta$ are the mean and the variance that are learned by a neural network, which is parameterized by $\theta$. The neural network is trained to predict the noise added at each timestep, allowing the model to reverse the noising process and recover an estimate of the original image from pure noise \cite{ddpm,ddpm_improved}. For the training objective of the reverse process, a variational lower bound $L$ on the data log-likelihood can be used, which can be simplified to a mean-squared error (MSE) between the predicted noise and the true noise added in each forward step \cite{ddpm,ddpm_improved}:

\begin{equation}
    L=\mathbb{E}_{t,x_0,\epsilon}[\parallel\epsilon-\epsilon_\theta(x_t,t)\parallel^2]
\end{equation}

where $\epsilon_\theta(x_t,t)$ is the neural network's prediction of the noise, and $\epsilon$ is the true noise added during the forward process.

\subsubsection{Generation Process}

Once the reverse process is learned, synthetic data generation can be achieved by starting from random Gaussian noise and iteratively applying the reverse diffusion steps \cite{ddpm,ddpm_improved}. The generation process begins by sampling a noisy image from a Gaussian distribution at timestep $T$ and then successively applying the reverse diffusion process to generate an image at each preceding timestep. This results in a sample $x_0$, which is a denoised image resembling the original training data, such as lung CT images in this case.

\subsection{Memory Efficient DDPM Image Generation}

The generation of images with diffusion models can use significant computational resources \cite{fastddpm,patchddm,latentmodel}. For this reason, Bieder et al. \cite{patchddm}, have developed a patch-wise approach to train a diffusion model to segment lesions in MRI scans. To be more memory efficient the model is trained on patches of the model input, thus the DDPM is trained on a smaller image size reducing spatial dimension and memory requirements. For this, the input is randomly cropped during training and the neural net only sees a part of the input each time. To guide the model additionally, the coordinates of the patch corresponding to the image are given as a condition, as well as the patch of the MRI image itself. Thus, the segmentation mask is noised for training, and in the generation approach the segmentation mask is generated out of Gaussian noise. We adapt this approach to generate whole lung CTs.

To be able to generate whole lung CTs with this approach we do not use the segmentation masks as the input that is noised. Instead, we noise the CT image patch and use the segmentation mask and the coordinates as a condition. The segmentation mask helps the model to additionally understand where to insert a nodule. The coordinate system gives the model an understanding of which part of the CT image it sees at all times. This way we can utilize the computational efficiency introduced in Bieder et al. \cite{patchddm} while generating the image instead of the segmentation mask. For sampling the images, we also sample the whole image directly like Bieder et al. \cite{patchddm}, not just patches. For this, we give the whole coordinate system and the whole segmentation mask to the trained DDPM.

\section{Experiments}

\subsection{Experiment Set-Up}

As mentioned before, the goal is to evaluate whether the generated images have a high utility in the real-world dataset. For all utility tests, we selected the nnU-Net pipeline proposed by Isensee et al. \cite{nnUnet}. The authors developed a self-configuring pipeline for medical image segmentation. This way one can add medical images into the pipeline and then data preprocessing and training of a U-Net is conducted. This U-Net can then segment test images. We decided to use this approach due to its ease of use and proven track record in image segmentation \cite{nnUnet, nnUnetrevised}. 

\begin{figure}[htbp]
\centerline{\includegraphics[scale=.18]{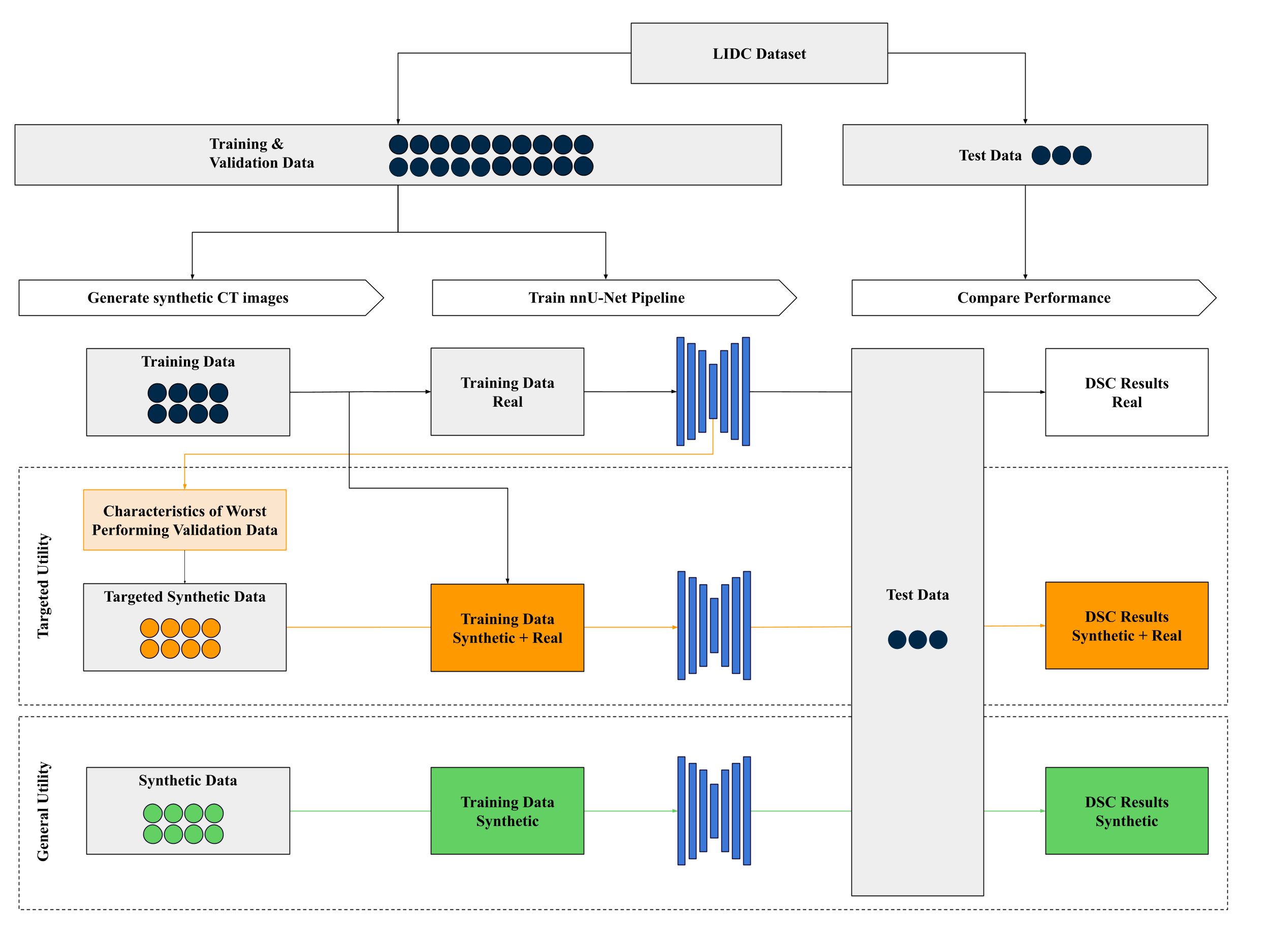}}
\caption{The experiment setup divides the LIDC-IDRI data into training, validation, and test data. From the training and validation data, a benchmark model is trained and the training data is used to train the diffusion model. Then a segmentation model is trained on synthetic data only (green). Additionally, based on the worst-performing validation data of the benchmark segmentation model, synthetic data is generated (orange) and a segmentation model is trained on real-world and the targeted synthetic data. All models are tested on the same real-world test set and DSCs are obtained.}
\label{fig:experiment-setup}
\end{figure}

As shown in Figure~\ref{fig:experiment-setup}, we divide the LIDC-IDRI dataset into training, validation, and test sets. The training set is used to train the mentioned nnU-Net and the DDPM. The validation set is only used to validate the nnU-Net and the test set is also only used to test the nnU-Net. As illustrated in the chart, we do a general and a targeted utility test. As the benchmark, the nnU-Net pipeline is trained with real-world LIDC-IDRI data only and tested on a fixed real-world LIDC-IDRI test set. For that, it is important to note that the nnU-Net pipeline includes data augmentation techniques. For the general utility test, the synthetic data is created only from the LIDC-IDRI segmentations of the training set. The nnU-Net pipeline is then trained with the generated synthetic data and also tested on the same fixed LIDC-IDRI test set. We then compare the DSC of both models created. The DSC tells us how much the predicted segmentation mask matches the segmentation mask of the test set. If the DSC of the model trained on real-world data and of the model trained on synthetic data is similar one can assume that the utility of the synthetic data is high. This is because the model seems to learn the right information from the synthetic data if it behaves similarly on a test set to the model trained on real-world data. 

For the targeted utility test, we investigate which validation images were predicted poorly and what kind of properties these images have, e.g. nodules close to the pleura. Our hypothesis is, that the model will improve if it sees more cases with similar properties. Thus, we then take the segmentation masks of these validation samples and generate more images. This way the newly generated data (called targeted synthetic data in this work) will have nodules of the same size, shape, and location. However, the model can vary the nodule type (solid, part-solid, or ground-glass) and inner structure of the lung. Then we train nnU-Net pipeline with the real-world LIDC-IDRI training data and the synthetic data with the specific segmentation masks. Then the DSC of the benchmark and the newly trained segmentation model are compared, to see if the information in the synthetic data improved the U-Net in its segmentation ability. 

Additionally, to ensure statistical significance, we train each model 5 times to determine a mean DSC and calculate the p-value between each mean DSC created. The DSC is calculated as follows:

\begin{equation}\label{eq:dice}
    DSC=\frac{2TP}{2TP+FP+FN}
\end{equation}

The True Positives (TP) in this case are nodule areas that are detected correctly, the False Positives (FP) are areas incorrectly detected as nodule areas and the False Negatives (FN) are nodule areas that were detected falsely. Thus this score shows how big the overlap between the pixels of the prediction and the pixels of the ground truth is while disregarding the True Negatives (TN). The TNs are non-nodule areas that are detected correctly.

\subsection{Benchmark for Experiments}

As mentioned before, we train the nnU-Net on real-world data as a benchmark for the experiments. For this, we use 553 images as training data, 142 images as validation data, and 138 images as test data. All of the images are cropped at the lung and the test-validation-training-split is done according to the patient IDs. The validation and test set are kept constant for all models trained. With this setup, we can reach a DSC of $0.4913$ with a standard deviation of $0.02733$. 

In general, segmenting small nodules is a challenging task as nodules are very small compared to the overall image and they can be quickly confused with other small white objects in the CT image by the model. Other segmentation approaches trained on the LIDC-IDRI dataset include Ali et al. \cite{AliRoiSeg} with a DSC of 81.1\% and Chen et al. \cite{ChenRoiSeg} with a DSC from 74.4 \% to 88.7\%. However, both approaches rely on determining a region of interest before segmenting the nodule. Halde et al. \cite{HALDEseg} and Akila Agnes et al. \cite{AkilaSeg} with a mean DSC of 97.15\% and 93.6 \% do not take all nodule types into account. On the other hand Sadremomtaz and Zadnorouzi \cite{SadremomtazSeg} with a DSC ranging from 56.92\% to 64.39\% segment the nodules on the whole lung and take all nodule types into account. Given that other approaches did not take the whole CT image or every nodule type into account and that segmenting nodules is an overall challenging task we determined our segmentation model to be a valid benchmark.

\subsection{General Utility Evaluation of Synthetic Data}

\begin{figure}[htbp]
\centerline{\includegraphics[scale=.07]{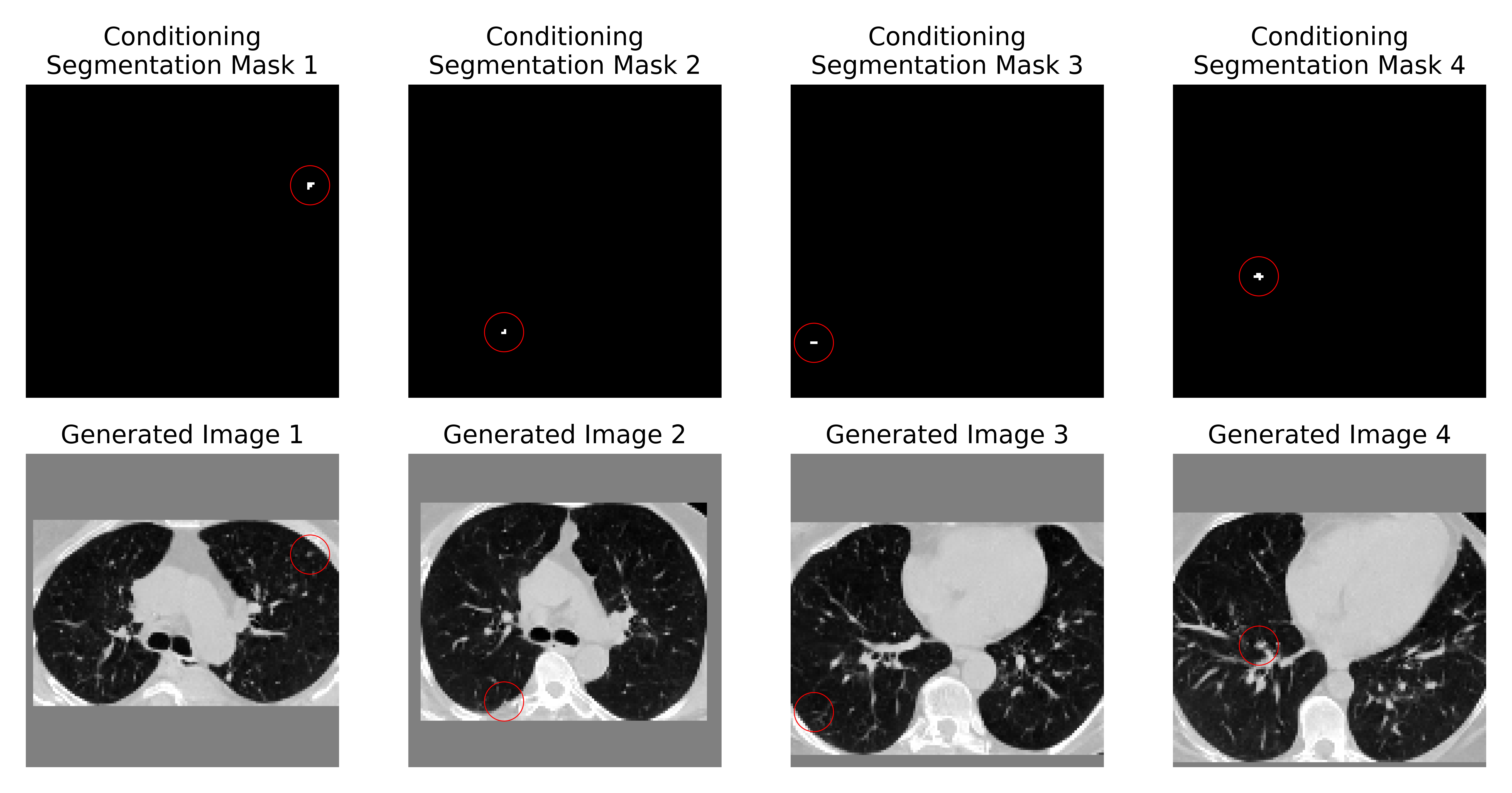}}
\caption{Generated images for a respective segmentation mask show that the model can hold the condition while generating an image.}
\label{fig:synthdata}
\end{figure}

In Figure~\ref{fig:synthdata} one can see some generated images. The first row has the conditional segmentation mask, as a basis for generation, and the next row has the generated images based on that mask. This showcases that the model can generate different lungs while still holding the condition of the conditional segmentation mask. However, a more general and comprehensive analysis lies in the utility tests.   

As mentioned, we train a nnU-Net with 553 generated CT images. As one can see in Table~\ref{tab:utility}, the mean DSC of the nnU-Net trained solely on real-world and solely on synthetic images are very close together, with the nnU-Net trained on real-world images having a mean DSC of $0.4913$ with standard deviation of $0.02733$ and the nnU-Net trained on solely synthetic data having a mean DSC of $0.5016$ with standard deviation of $0.0206$. The mean DSC of the synthetic segmentation model lies within the standard deviation of the mean DSC of the real-world segmentation model. Thus, this indicates that the segmentation model can learn similar information from the synthetic data as from the real-world data, indicating high utility for the synthetic data. 

\begin{table}
 
  \centering
  \begin{tabular}{lll}
    \toprule
    Training Data     & Mean Test Dice Score  & $p$-Value\\
    \midrule
    Real & $0.4913 \pm 0.02733$  &  -  \\
    Synthetic     & $0.5016 \pm 0.0206$ & -   \\ 
    Real + Synthetic     &  $0.5418 \pm 0.03015$ & $0.024335$ \\ 
    \bottomrule
  \end{tabular}
  \caption{The DSC of the Real segmentation model and the segmentation model trained on only synthetic data show a similar performance for both. The DSC of the model trained on real and targeted synthetic data shows an improvement in model performance. This indicates that the synthetic data is of high quality.}
  \label{tab:utility}
\end{table}

\subsection{Targeted Utility Evaluation of Synthetic Data}

As a first step, we evaluate our benchmark segmentation model and filter out the $59$ worst-performing validation samples. The $59$ worst-performing samples are all of the samples below the mean baseline DSC. With the segmentation mask of these validation samples, we generate synthetic images. As explained before, we then train a segmentation model with the real-world as well as targeted synthetic data. As one can see in Table~\ref{tab:utility}, training the model with real-world and synthetic data (named the Real and Real+Synthetic model in this work) improved the performance on the test set the DSC from $0.4913$ to $0.5418$. Taking the standard deviation into account the $p$-value for the improvement in the test set is $0.024335$. The improvement is statistically significant, as a $p$-value of $0.05$ and below is assumed to be statistically significant. 

\begin{figure}[htbp]
\centerline{\includegraphics[scale=.044]{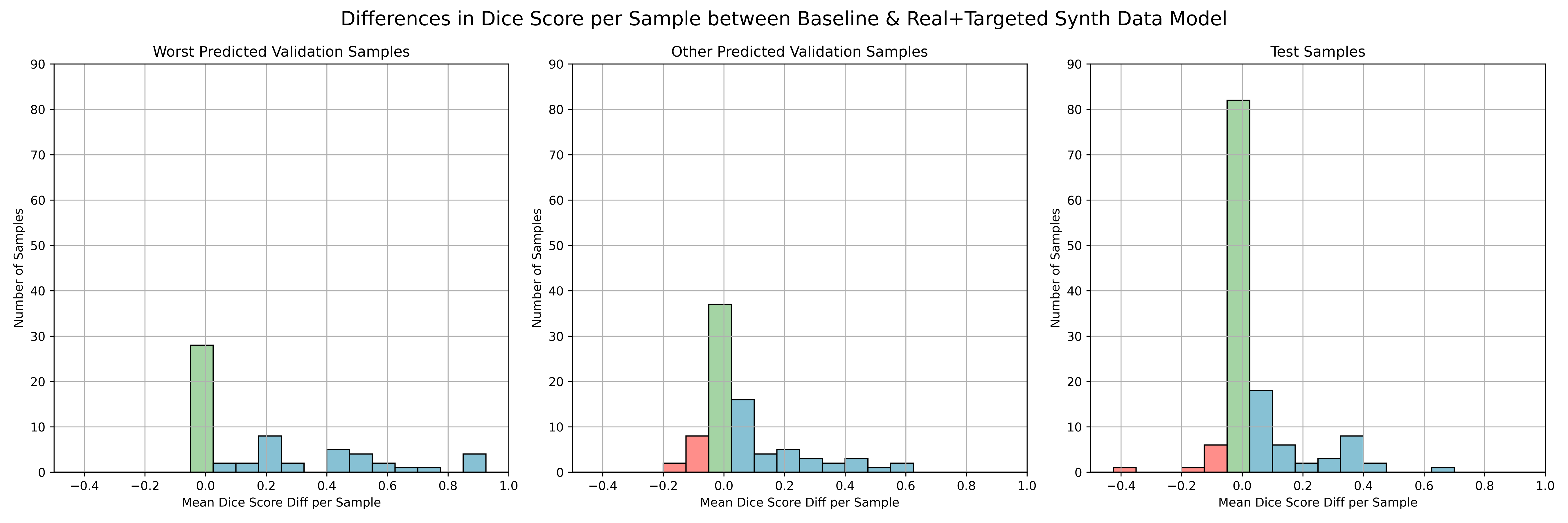}}
\caption{Difference in DSC between the Real and Real+Synthetic segmentations for the worst predicted validation samples, the other validation samples, and the test samples. The worst predicted validation CTs could be improved or did not change in value. For the other validation samples and the test samples the majority of the samples did not change or improve, showing an overall performance improvement of the classifier.}
\label{fig:dicedifference}
\end{figure}

\begin{figure}[htbp]
    \centering
    \includegraphics[width=0.9\linewidth]{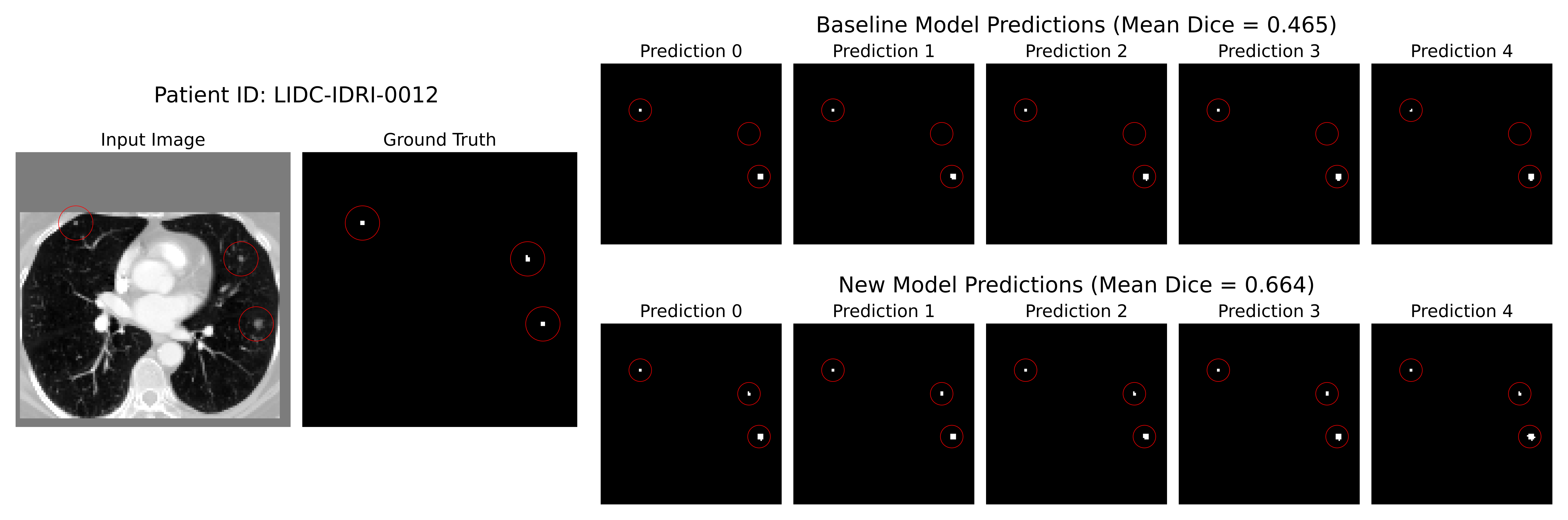}
    \caption{Validation CT Image with three nodules and the $5$ predictions of the Real and Real+Synthetic segmentation model. The Real model was only able to segment two. After adding the synthetic data the model could predict all 3.}
    \label{fig:val12}
\end{figure}

\begin{figure}[htbp]
    \centering
    \includegraphics[width=0.9\linewidth]{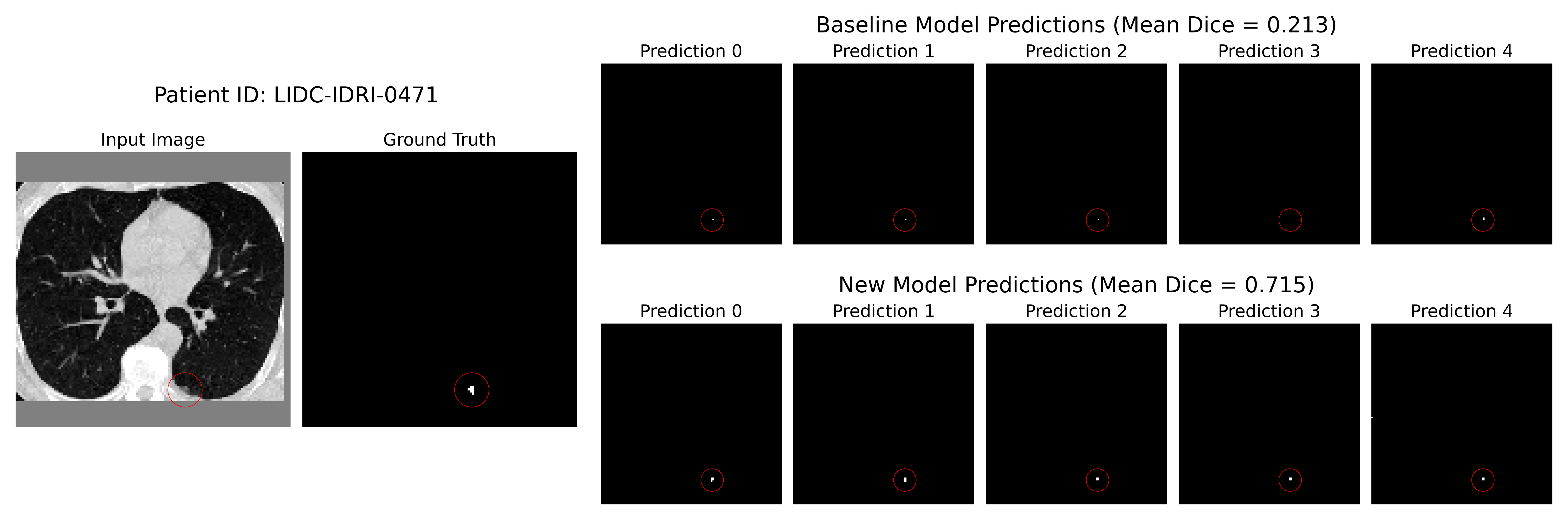}
    \caption{Validation CT Image with a nodule attached to the pleura and the $5$ predictions of the Real and Real+Synthetic segmentation model. The Real model was not able to distinguish nodule and pleura well. After adding the synthetic data the segmentation improved.}
    \label{fig:val505}
\end{figure}

\begin{figure}[htbp]
    \centering
    \includegraphics[width=0.9\linewidth]{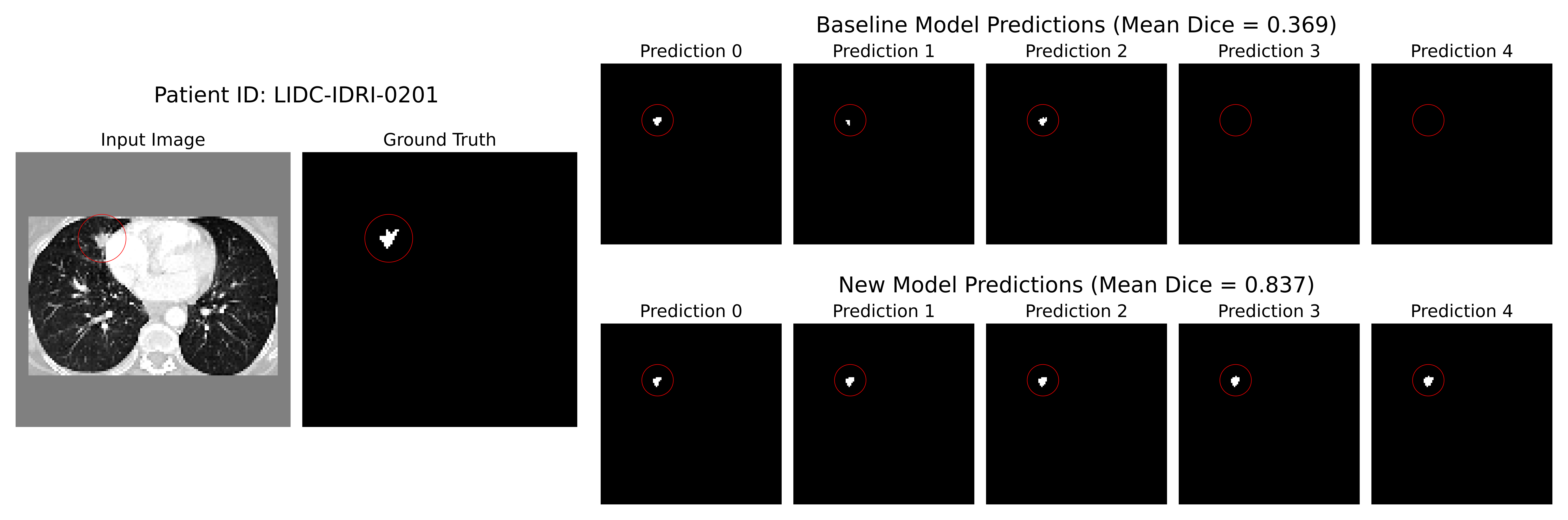}
    \caption{Test CT Image with a nodule attached to an organ and the $5$ predictions of the Real and Real+Synthetic segmentation model. The Real model was not able to distinguish nodule and organ well. After adding the synthetic data the segmentation improved.}
    \label{fig:test201}
\end{figure}

\begin{figure}[htbp]
    \centering
    \includegraphics[width=0.9\linewidth]{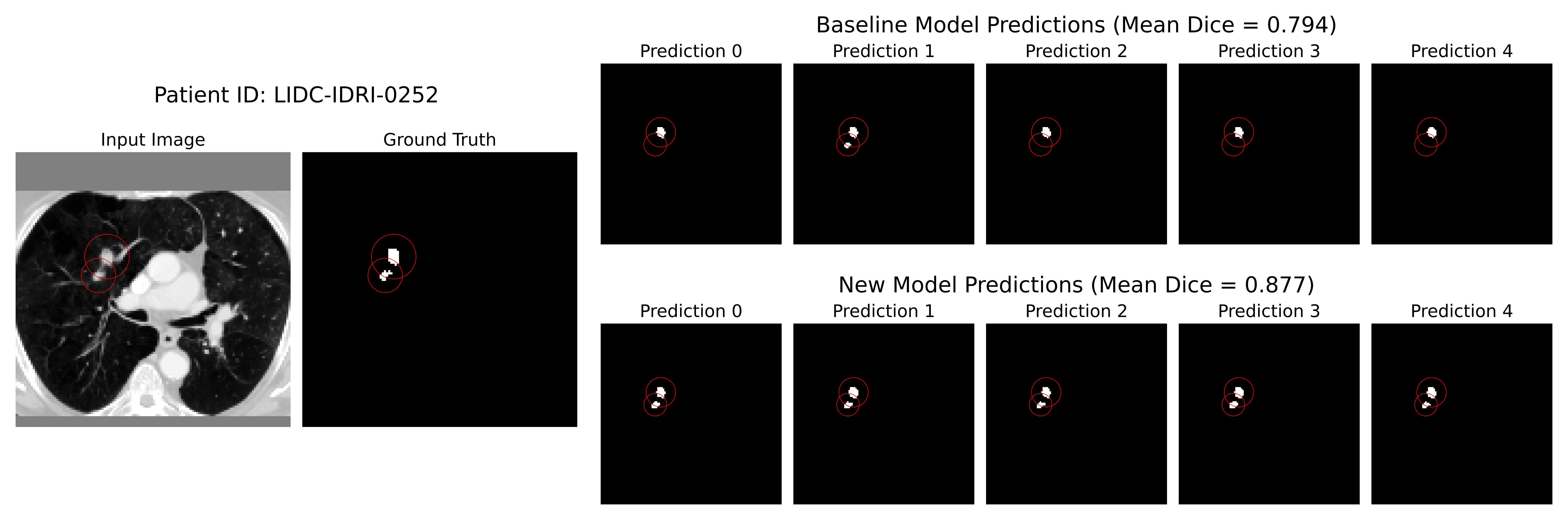}
    \caption{Test CT Image with two nodules close together and the $5$ predictions of the Real and Real+Synthetic segmentation model. The Real model had trouble segmenting both nodules. After adding the synthetic data both nodules were segmented in each prediction.}
    \label{fig:test252}
\end{figure}

\begin{figure}[htbp]
    \centering
    \includegraphics[width=0.9\linewidth]{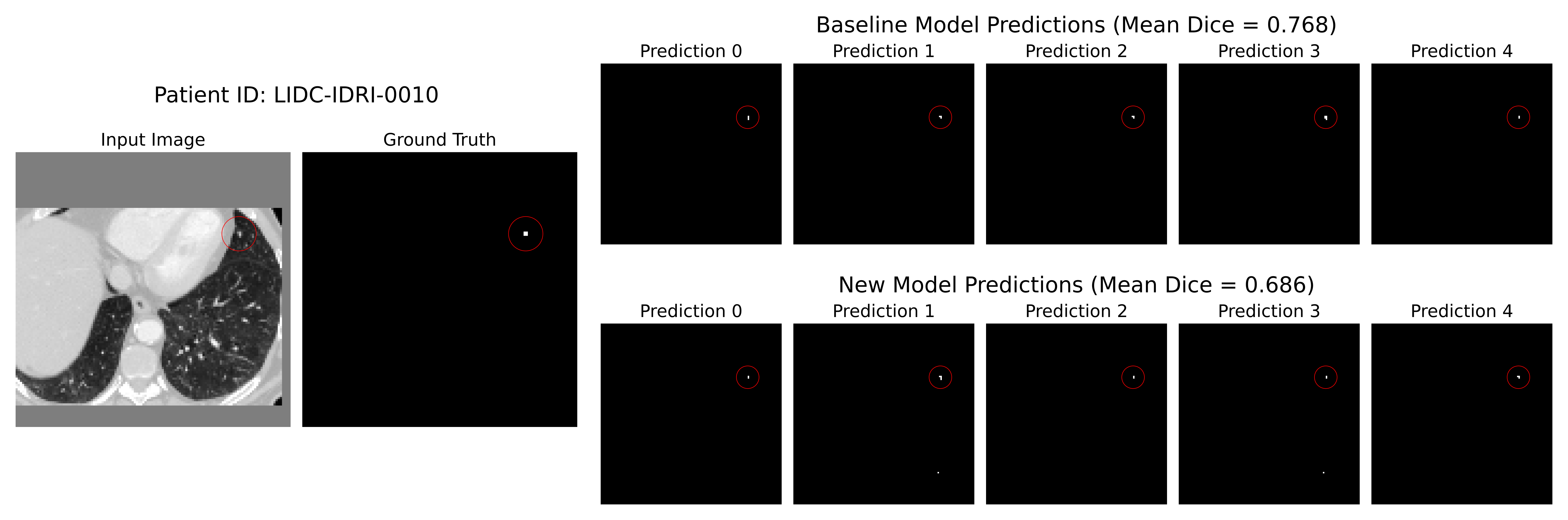}
    \caption{Test CT Image with one small nodule and the $5$ predictions of the Real and Real+Synthetic segmentation model. The Real model was only segmenting one nodule. After adding the synthetic data the model segmented two tumors.}
    \label{fig:test10}
\end{figure}

Looking at Figure~\ref{fig:dicedifference}, we can observe how the DSC has changed for 3 subgroups (the $59$ worst predicted validation samples, the other validation samples, and the test samples) compared to the benchmark model. For the Real+Synthetic model, the worst-performing validation samples have improved or produced the same DSC as before. This shows that the synthetic data inhibits the information it was supposed to reflect: the properties of the worst predicted validation samples. Figure~\ref{fig:val12} shows an example of this, where the model was able to segment all 3 tumors after the synthetic data was added. Another example, where a nodule attached to the pleura was segmented more accurately can be observed in Figure~\ref{fig:val505}. Looking at the other validation samples, the majority has improved the DSC or it did not change at all, some samples have become worse in the process. For the test samples, most images did not change in DSC, and more predictions improved than worsened in the process. An example of an improved segmentation can be seen in Figure~\ref{fig:test201}, where the segmentation of a nodule attached to an organ was improved. In Figure~\ref{fig:test252}, two nodules close together were segmented each time by the Real+Synthetic model while the Real model struggled to do so. As we added more edge cases through the validation dataset like images with more nodules than average, like in Figure~\ref{fig:val12}, it also happened that the Real+Synthetic model over-segmented some images like in Figure~\ref{fig:test10}. The overall test DSC improvement, the histogram of test DSCs, and other validation DSCs showed that we did not overfit on the worst predicted validation samples. However, a more comprehensive analysis, of properties that the model still needed to learn, would be beneficial to avoid cases like the over-segmentation. This way one could do a more targeted generation and make sure that even fewer samples worsen in DSC. 

\section{Conclusion}
In this study, we addressed the challenge of data scarcity in medical imaging by developing a memory-efficient patch-wise DDPM method to generate synthetic CT scans with lung nodule segmentation. Our experiments showed that training on synthetic data alone resulted in a competitive DSC (0.5016) compared to models trained on real-world data (0.4913). This highlighted the utility of synthetic images in segmentation tasks. Moreover, augmenting real-world data with synthetic images improved the overall performance, achieving a DSC of 0.5418 on the test set, showing the added value of synthetic data in improving model segmentation. The model successfully inserted a nodule according to the conditional segmentation mask for the generated synthetic data. This shows the model’s capability to generate contextually appropriate synthetic images.

However, despite the overall improvement in performance, we observed that in some cases the DSC deteriorated. For example, particularly with more challenging samples, the model detected nodules that were absent. This suggests that while synthetic data enhances model performance, a more thoughtful generation strategy is needed to avoid introducing erroneous patterns, especially when dealing with complex anatomical structures like lung nodules.

Overall, our findings demonstrate the potential of synthetic data to augment medical image datasets and improve segmentation models, but they also emphasize the importance of careful consideration when generating synthetic data to ensure reliable model predictions.

\section*{Acknowledgments}
This work was supported in part by Malika Sanhinova, who assisted with running experiments for the study. We also acknowledge the MONAI project for their open-source code, which was used in portions of this research.

\bibliographystyle{IEEEtran}
\bibliography{references}

\end{document}